\documentclass[aps,
  prb,
  groupedaddress,
  twocolumn,
  amsmath,
  floatfix
]{revtex4-2}

\makeatletter
\newif\iftwocol
\@ifclasswith{revtex4-2}{twocolumn}{\twocoltrue}{\twocolfalse}
\makeatother

\iftwocol
  \newcommand\iwm{.97}
  \newcommand\rmw{.49}
\else
  \newcommand\iwm{.7}
  \newcommand\rmw{.97}
\fi

\usepackage[utf8]{inputenc}
\usepackage[caption=false]{subfig}
\usepackage[english]{babel}
\usepackage{graphicx}
\usepackage{xcolor}
\usepackage{siunitx}
\usepackage[nolist]{acronym}
\usepackage{comment}
\usepackage{soul}
\usepackage{xspace}
\usepackage[hidelinks]{hyperref}

\graphicspath{{../img/}{./}}

\usepackage{mathtools}

\newcommand\figref[1]{Fig.\,\ref{fig:#1}}
\newcommand\subfigref[1]{Fig.\,\ref{subfig:#1}}

\newcommand\eqnref[1]{(\ref{eqn:#1})}

\usepackage{etoolbox}
\robustify{\subref}

\usepackage{bbold}
\newcommand\ident{{\ensuremath\mathbb{1}}}

\newcommand\mat[1]{\ensuremath{\mathbf{#1}}}
\newcommand\op[1]{\ensuremath{\mathbf{#1}}}
\renewcommand\vec[1]{\ensuremath{\mathbf{#1}}}
\newcommand\matd[1]{\ensuremath{\mat{#1}^\dagger}}
\newcommand\opd[1]{\ensuremath{\op{#1}^\dagger}}

\newcommand*\diff{\mathop{}\!\mathrm{d}}

\newcommand\s[1]{\ensuremath{s_{#1}}}
\newcommand\m[1]{\ensuremath{\Gamma_{#1}}}
\newcommand\p[1]{\ensuremath{\phi_{#1}}}
\newcommand\w[1]{\ensuremath{\tau_{#1}}}

\newcommand\eig[1]{\ensuremath{\acs{eig}\left(#1\right)}}

\newcommand\Tsym{\mathcal{T}}
\newcommand\Psym{\mathcal{P}} 
\newcommand\Top{\op{\Tsym}}
\newcommand\Pop{\op{\Psym}}

\newcommand\Topd{\opd{\Tsym}}
\newcommand\Popd{\opd{\Psym}}
\newcommand\applyT[1]{\Top\,#1\,\Topd}
\newcommand\applyP[1]{\Pop\,#1\,\Popd}
\newcommand\applyPT[1]{\Pop\Top\,#1\,\Topd\Popd}

\makeatletter
\newcommand{\acms}{\protect\@acms}%
\newcommand{\@acms}[1]{%
    \expandafter\ifx\csname ac@#1\endcsname\AC@used
        \ensuremath{\acs{#1}}%
    \else%
        \acl{#1} \ensuremath{\acs{#1}}\acused{#1}%
    \fi%
}%
\newcommand{\acm}{\protect\@acm}%
\newcommand{\@acm}[1]{%
    \expandafter\ifx\csname ac@#1\endcsname\AC@used
        \ensuremath{\acs{#1}}%
    \else%
        the \acl{#1} \ensuremath{\acs{#1}}\acused{#1}%
    \fi%
}%
\newcommand{\acmp}{\protect\@acmp}%
\newcommand{\@acmp}[1]{%
    \expandafter\ifx\csname ac@#1\endcsname\AC@used
        \ensuremath{\acs{#1}}%
    \else%
        the \aclp{#1} \ensuremath{\acs{#1}}\acused{#1}%
    \fi%
}%
\newcommand{\Acm}{\protect\@Acm}%
\newcommand{\@Acm}[1]{%
    \expandafter\ifx\csname ac@#1\endcsname\AC@used
        \ensuremath{\acs{#1}}%
    \else%
        The \acl{#1} \ensuremath{\acs{#1}}\acused{#1}%
    \fi%
}%
\newcommand{\Acmp}{\protect\@Acmp}%
\newcommand{\@Acmp}[1]{%
    \expandafter\ifx\csname ac@#1\endcsname\AC@used
        \ensuremath{\acs{#1}}%
    \else%
        The \aclp{#1} \ensuremath{\acs{#1}}\acused{#1}%
    \fi%
}%
\newcommand{\acml}[1]{the \acl{#1} \ensuremath{\acs{#1}}\acused{#1}}

\newcommand{\acmsl}[1]{\acl{#1} \ensuremath{\acs{#1}}\acused{#1}}

\makeatother

\begin{document}

\preprint{Ver. 12}

\title{Non-Reciprocity of the Wave Packet Scattering Delay in Ballistic Two-Terminal Devices}

\author{P. Bredol}
\affiliation{Max Planck Institute for Solid State Research,
             70569 Stuttgart, Germany}

\date{\today}

\begin{abstract}
In the linear regime, transport properties of ballistic two-terminal devices are generally considered to be independent of the direction of the current. This two-terminal reciprocity applies to both the electron transmission and reflection probabilities. However, it does not apply to the scattering delays of wave packets. Indeed, \emph{four} different time delays describe the transmission and reflection processes of wave packets arriving from the two terminals, respectively. Unlike the probabilities, these delays are direction dependent if the channel exchange symmetry of the scattering matrix is broken.
\end{abstract}


\maketitle

\begin{acronym}
  \acro{ab}[AB]{Aharonov-Bohm}
  \acro{S}[\mat{S}]{scattering matrix}
  \acro{Q}[\mat{Q}]{Wigner-Smith time delay matrix}
  \acro{Sbs}[$\mat{S}_{BS}$]{beam splitter scattering matrix}
  \acro{k}[$k$]{wave vector}
  \acro{Sd}[\matd{S}]{scattering matrix}
  \acro{tro}[\Top]{time-reversal operator}
  \acro{ceo}[\Pop]{channel exchange operator}
  \acro{w}[$w$]{width}
  \acro{Vl}[$V_l$]{left side potential height}
  \acro{Vr}[$V_r$]{right side potential height}
  \acro{flux}[$\Phi$]{enclosed magnetic flux}
  \acro{flux0}[$\Phi_0$]{magnetic flux quantum}
  \acro{l}[$l$]{arm length}
  \acro{dl}[$\Delta l$]{arm length difference}
  \acro{dens}[$n$]{probability amplitude}
  \acro{eig}[$\Phi$]{eigenfunction}
  \acro{dphi}[$\Delta\phi$]{phase difference}
  \acro{k0}[$k_0$]{central momentum}
  \acro{ks}[$\sigma$]{momentum spread}
  \acro{x0}[$x_0$]{position}
  \acro{x0l}[$x^L_0$]{left side position}
  \acro{x0r}[$x^R_0$]{right side position}
\end{acronym}

\newcommand\natunits{${\hbar=q=2m=1}$\xspace}

\newcommand\valueW{5}
\newcommand\valueVl{0}
\newcommand\valueVr{5}
\newcommand\valueFlux{\frac{\pi}{2}}
\newcommand\valueL{5}
\newcommand\valueDl{\frac{\pi}{4}}
\newcommand\valuekO{\pm2}
\newcommand\valueks{0.4}
\newcommand\valuexOl{-15}
\newcommand\valuexOr{20}

\section{Introduction}
Ballistic two-terminal electron devices are known to have reciprocal transport properties in the linear response regime: the transmission of the electrons through these devices is independent of their travel direction. This seems to be obvious, because the transmission and reflection probabilities have to be independent of the direction of the current flow due to the unitarity of the underlying scattering matrix. However, characterizing a device only by the transmission and reflection probabilities neglects all temporal aspects of the electron transport process. These aspects are captured by the phases of the scattering coefficients, which do not have to be reciprocal.

The wave packet scattering delay is given by the derivative of the complex phase of a scattering coefficient with respect to energy and quantifies the duration of the reflection or transmission process of a wave packet interacting with a scatterer \cite{lower-limit-for-the}. For a system with two scattering channels, four of such time delays exist, corresponding to the four entries of the scattering matrix. Three of these four delays can be chosen independently by design of the scattering region, which allows non-reciprocal delays to be implemented. Interestingly, it is the device symmetry, which determines the (non-)reciprocity of the scattering delays, as I present in this contribution.

The second well established way of characterizing the lifetimes of scattering states is given by the \acl{Q} \cite{lifet-matri-in-colli}. The $n$-th diagonal element of the \acl{Q} corresponds to the sum of the scattering delays for wave packets starting in channel $n$ weighted by the respective transmission (reflection) probabilities. Here, I focus on the wave packet scattering delay approach, because it distinguishes between transmission and reflection more straightforwardly than the delays and lifetimes extracted from the \acl{Q}. The revealed asymmetries of the wave packet scattering delays are obviously inherited by the Wigner-Smith time delays.

In optical setups, the wave packet scattering delays can be as large as nanoseconds. Thus, in the optical case the direct measurement of these delays is technologically feasible and was presented in multiple experiments \cite{direc-measu-of-the1,direc-measu-of-the2}. The delays of electrons traveling through mesoscopic structures are on the order of picoseconds or femtoseconds, however. Measuring the dynamics of electrons in semiconductor systems therefore requires single or few electron devices and picosecond resolution, which was achieved in a recent experiment \cite{picos-coher-elect}.

In the following, I will consider two-channel systems, for which the constraints imposed by the unitarity of the \acl{S} are particularly strict and enforce full reciprocity of the scattering probabilities. Systems with a larger number of channels feature more entries of the \acl{S}, which relaxes these constraints. For the two-channel case it will be shown that the \emph{transmission} delays are non-reciprocal in cases where the scattering matrix breaks time-reversal symmetry. The \emph{reflection} delays are non-reciprocal if the scattering matrix is asymmetric with respect to a simultaneous time reversal and exchange of scattering channels. The non-reciprocity of the wave packet scattering delays is demonstrated using two analytically solvable examples: an asymmetric potential barrier and an asymmetric \ac{ab} interferometer. The peculiarities of the wave packet transport in such interferometers was previously presented in \cite{nonre-inter-for}. The consequences for both time-dependent and stationary wave functions are presented, and implications for quantum information technology and nanoelectronics are discussed.

\section{Symmetries of the Scattering Matrix}

To determine the symmetry properties of the wave packet scattering delays of a system, a three dimensional scattering region and two one-dimensional semi-infinite channels are assumed. The position of the scattering region is given by ${0\leq x\leq R}$. The first channel extends in negative $x$ direction (${x<0}$), and the second one extends in positive $x$ direction (${x>R}$). Outside the scattering region the eigenfunctions \eig{\vec{r},k} are given by plane waves:
\begin{align}
  \eig{\vec{r},k} = \begin{cases}
    \left(A_1e^{ikx} + B_1e^{-ikx}\right)\phantom{A}\\
            \hfill\cdot\delta(y)\delta(z)
                &\text{for } x < 0,\\[.5em]
    \left(A_2e^{-ikx} + B_2e^{ikx}\right)\phantom{A}\\
            \hfill\cdot\delta(y)\delta(z)
                &\text{for } x > R,\\[.5em]
    \tilde\Phi\left(\vec{r},k\right) &\text{else},
  \end{cases}
  \label{eqn:eigfunc}
\end{align}
with the delta distribution $\delta$, ${\vec{r}=(x,y,z)}$, complex amplitudes $A_1$, $A_2$, $B_1$, $B_2$ and \acm{k}. $\tilde\Phi$ denotes the wave function inside the scattering region. \Acm{S} transforms the pair of incoming wave amplitudes into the pair of outgoing wave amplitudes ${(B_1,B_2)^T=\acs{S}(A_1,A_2)^T}$ and reads:
\begin{align}
  \acs{S} = \begin{pmatrix}
    \s{11} & \s{12} \\ \s{21} & \s{22}
  \end{pmatrix}
  = \begin{pmatrix}
    \m{11}e^{i\p{11}} & \m{12}e^{i\p{12}} \\
    \m{21}e^{i\p{21}} & \m{22}e^{i\p{22}}
  \end{pmatrix}.
\end{align}

To satisfy probability conservation, ${\matd{S}\acs{S}=\ident}$ must hold. This condition requires ${\m{11}^2+\m{12}^2=1}$, ${\m{12}=\m{21}}$ and ${\m{11}=\m{22}\eqqcolon\m{}}$. The scattering matrix then reads
\begin{align}
  \acs{S} = \begin{pmatrix}
    \m{}e^{i\p{11}} & \sqrt{1-\m{}^2}e^{i\p{12}} \\
    \sqrt{1-\m{}^2}e^{i\p{21}} & \m{}e^{i\p{22}}
  \end{pmatrix}.
\end{align}

 \s{21} and \s{12} (\s{11} and \s{22}) differ only by a phase factor, the transmission probability from channel 1 to channel 2 (${|\s{21}|^2}$) is identical to the transmission probability from channel 2 to channel 1 (${|\s{12}|^2}$). The reflection probabilities show an analogous behavior. Thus, all four probabilities are invariant under the exchange of scattering channels, even if the Hamiltonian of the scattering region does not bear this symmetry.

The consequences of this strict reciprocity are remarkable: \ac{ab} conductance oscillations in two-terminal interferometers are always symmetric with respect to zero magnetic field. Even if the spatial symmetry is broken by the geometry or by applied gate potentials, it is not possible to tune the phase of the oscillation pattern continuously. However, adding further scattering channels removes this so-called phase rigidity, which makes arbitrary oscillation phases possible \cite{contr-of-the-trans}.

As seen above, the scattering probabilies are fully constrained by the unitarity condition. For the scattering phases $\p{ij}$, however, unitarity implies only one condition, namely
\begin{align}
  \p{11}+\p{22}=\p{21}+\p{12}+\pi.
\end{align}
The three remaining degrees of freedom are controlled by two system symmetries: the symmetry with respect to \acm{ceo} and to \acm{tro}. Outside the scattering region, the effect of these two operators on the system eigenfunctions as defined in \eqnref{eigfunc} can be understood without knowledge of the scattering region. If \acs{ceo} is applied to an eigenfunction, it exchanges $A_1$ with $A_2$ and $B_1$ with $B_2$. Applying \acs{tro} replaces $k$ with $-k$ and conjugates all complex amplitudes $A_1$, $A_2$, $B_1$ and $B_2$. The effect of the symmetry operations on \acs{S} is found by applying the operators to the equation
${(B_1,B_2)^T=\acs{S}(A_1,A_2)^T}$. Applying \acs{tro}, we obtain
\begin{align}
  \begin{pmatrix}A_1^*\\A_2^*\end{pmatrix}
    = \applyT{\acs{S}}
    \begin{pmatrix}B_1^*\\B_2^*\end{pmatrix}.
\end{align}
The time-reversed equation shows that \applyT{\acs{S}} is the conjugated inverse matrix $(\mat{S}^{-1})^*$. Owing to ${\acs{S}^{-1}=\acs{S}^\dagger}$, applying \acs{tro} to \acs{S} is equivalent to transposing \acs{S}. Applying \acs{ceo} to ${(B_1,B_2)^T=\acs{S}(A_1,A_2)^T}$, we obtain
\begin{align}
  \begin{pmatrix}B_2\\B_1\end{pmatrix}
    = \applyP{\acs{S}}
    \begin{pmatrix}A_2\\A_1\end{pmatrix}.
\end{align}
The channel exchange transposes \acs{S} and exchanges the diagonal elements. Thus, the combined operation \acs{ceo}\acs{tro} exchanges only the diagonal elements of \s{11} and \s{22}. The following relations hold for the phases of the scattering coefficients:
\begin{alignat}{3}
  \acs{S}=\applyP{&\acs{S}} \quad&&\Leftrightarrow\quad\p{11}=\p{22}\text{ and }\p{12}=\p{21}
  \label{eqn:psym},\\
  \acs{S}=\applyT{&\acs{S}} \quad&&\Leftrightarrow\quad\p{12}=\p{21}
  \label{eqn:tsym},\\
  \acs{S}=\applyPT{&\acs{S}}\quad&&\Leftrightarrow\quad\p{11}=\p{22}
  \label{eqn:ptsym}.
\end{alignat}

If \eqnref{psym} holds, full reciprocity follows. If \eqnref{psym} is broken, but \eqnref{tsym} holds, \eqnref{ptsym} is broken as well. Consequently, the transmission phases are reciprocal but the reflection phases are not. Vice versa, if \eqnref{psym} and \eqnref{tsym} are broken, but \eqnref{ptsym} holds, the reflection phases are reciprocal but the transmission phases are not.

In the following, two analytically solvable example systems are presented, each of which fulfills one of the symmetries \eqnref{tsym} or \eqnref{ptsym}, but break the other one. The results are obtained by solving the one-dimensional Sch\"odinger equation for a free particle with mass $m$ and charge $q$ in presence of an electromagnetic vector potential \vec{A}. In natural units \natunits, the Hamiltonian reads
\begin{align}
  \op{H} = (-i\partial_x-\vec{A})^2.
  \label{eqn:hamiltonian}
\end{align}

\begin{figure}
  \centering
  \subfloat[\label{subfig:lrat-sys}]{\begin{minipage}{\linewidth}
    \includegraphics{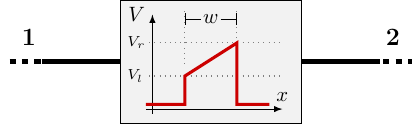}
  \end{minipage}}\newline
  \subfloat[\label{subfig:lrat-scatter}]{
    \centering
    \includegraphics[width=\iwm\linewidth]{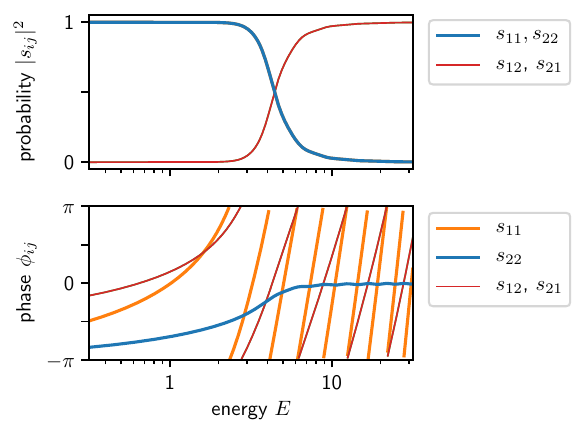}
  }
  \caption{\label{fig:lrat}\subref{subfig:lrat-sys} Asymmetric potential barrier embedded in an infinite waveguide. Scattering channel 1 and 2 correspond to the waveguide leaving to the left and right, respectively. The potential is characterized by its \acmsl{w}, \acml{Vl} and \acml{Vr}. \subref{subfig:lrat-scatter} Polar representation of the scattering coefficients calculated as a function of energy for $\acs{w}=\valueW$, $\acs{Vl}=\valueVl$ and $\acs{Vr}=\valueVr$ in natural units \natunits.}
\end{figure}

The potential barrier shown in \figref{lrat} is time-reversal symmetric but not invariant under channel exchange. It fulfills \eqnref{tsym} but breaks \eqnref{ptsym}. Figure \ref{subfig:lrat-scatter} shows the resulting scattering coefficients as a function of energy. As expected, both transmission coefficients are identical and the reflection coefficients differ by a phase factor. For energies much lower than the barrier height, the transmission probability vanishes, whereas for energies much higher than the barrier height, transparency is approached. As dictated by unitarity, the transmission and reflection probabilities are identical for forward and reverse direction.

\begin{figure}
  \subfloat[\label{subfig:trat-sys}]{\begin{minipage}{\linewidth}
    \includegraphics{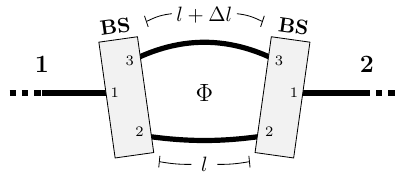}
  \end{minipage}}\\
  \subfloat[\label{subfig:trat-scatter}]{
    \includegraphics[width=\iwm\linewidth]{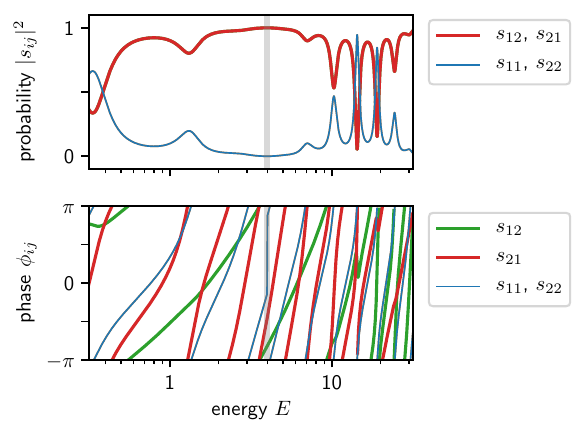}
  }
  \caption{\label{fig:trat}\subref{subfig:trat-sys} Asymmetric \ac{ab} interferometer composed of four waveguides (black lines) and two beam splitters (gray boxes). The small numbers 1--3 denote the respective channel indices of \acm{Sbs}. The composed system has two channels (bold numbers), which correspond to the waveguides leaving to the left and right, respectively. The interferometer is characterized by \acml{flux} and the arm lengths \acs{l} and ${\acs{l}+\acs{dl}}$. \subref{subfig:trat-scatter} Polar representation of the scattering coefficients calculated as a function of energy for $\acs{l}=\valueL$, $\acs{dl}=\valueDl$ and $\acs{flux}=\valueFlux$ in natural units \natunits. Transparency is achieved for $E=4$ (gray line). The phase jump of $\pi$ in the reflection phase is in accordance with the corresponding zero in the reflection probability.}
\end{figure}

The asymmetric \ac{ab} interferometer shown in \figref{trat} consists of four one-dimensional waveguides connected by two beam splitters. The eigenfunctions in the waveguides are described by plane waves, and the two beam splitters are characterized by the energy-independent three-channel scattering matrix
\begin{align}
  \acs{Sbs} = \begin{pmatrix}
    0 & \frac{1}{\sqrt{2}} & \frac{1}{\sqrt{2}} \\
    \frac{1}{\sqrt{2}} & -\frac{1}{2} & \frac{1}{2} \\
    \frac{1}{\sqrt{2}} & \frac{1}{2} & -\frac{1}{2}
  \end{pmatrix}.
\end{align}

As depicted in \figref{trat}, channel 1 of each beam splitter is connected to one of the infinite waveguides leaving to the left and right. Channels 2 and 3 of the beam splitters are connected to the lower and upper interferometer arm, respectively. The asymmetric \ac{ab} interferometer is neither time-reversal symmetric nor symmetric with respect to channel exchange. Both symmetry operations are equivalent to the reversal of the enclosed magnetic flux. Thus, the combined operation \acs{ceo}\acs{tro} leaves the system unchanged, and the interferometer fulfills \eqnref{ptsym} but breaks \eqnref{tsym}. As \subfigref{trat-scatter} confirms, both reflection coefficients are identical, and the transmission coefficients differ by a phase factor. The transmission and reflection probabilities show numerous resonances due to the energy dependence of the phase difference acquired between the two arms
\begin{align}
  \acs{dphi}={\acs{dl}\acs{k}\pm2\pi\acs{flux}/\acs{flux0}},
  \label{eqn:dphi}
\end{align}
with the \acl{flux0} ${\acs{flux0}=h/e=2\pi}$ in the chosen natural units. The plus and minus sign correspond to the ${1\rightarrow 2}$ and ${2\rightarrow 1}$ directions, respectively. If \acm{dphi} is a multiple of $2\pi$ and with \acs{Sbs} as defined above, full transparency follows. Using ${\acs{flux0}=2\pi}$, $E=k^2$ and the system parameters ${\acs{dl}=\valueDl}$ and ${\acs{flux}=\valueFlux}$, equation \eqnref{dphi} shows that transparency is achieved for $E=4$. In all other cases, the behavior cannot be easily understood by tracing individual paths and their accumulated \aclp{dphi} \acs{dphi} because many partial reflections at the beam splitters and a large number of round trips in the interferometer may occur along the paths.

\section{Non-Reciprocity of the Wave Packet Scattering Delay}

After the previous section demonstrated the effects of the system symmetries on the scattering coefficients and their non-reciprocal phases, in this section the resulting non-reciprocal delays are presented. The wave packet scattering delay is defined as the derivative of the scattering phase with respect to energy \cite{lower-limit-for-the},
\begin{align}
  \w{ij} = \hbar\frac{\partial\p{ij}}{\partial E}.
  \label{eqn:delay}
\end{align}
To understand this definition, it is instructive to study the undisturbed motion of a Gaussian wave packet first. According to the Hamiltonian \eqnref{hamiltonian}, the temporal evolution of such a wave packet in an infinite one-dimensional waveguide reads
\begin{align}
  \psi(k,t) = A\exp\left(-\frac{\left(k-\acs{k0}\right)^2}{2\acs{ks}^2}
                  - ik\left[\acs{x0} + kt\right]\right),
\end{align}
where $k=\pm\sqrt{E}$ is the momentum. The wave packet is centered around \acs{k0} with width \acs{ks} in momentum space; in position space it is centered at ${\acs{x0} + kt}$. $A$ is the normalization constant. After transmission or reflection, as described by one of the coefficients ${\s{ij}=\m{ij}e^{i\p{ij}}}$, the wave function is given by
\begin{align}
  &\s{ij}(k)\psi(k,t) = \m{ij}(k) \nonumber\\
  &~\cdot A\exp\left(-\frac{\left(k-k_0\right)^2}{2\acs{ks}^2}
                  - ik\left[x_0 + k\left(t - \frac{\p{ij}(k)}{k^2}\right)\right]\right).
\end{align}

\begin{figure}
  \centering
  \subfloat[\label{subfig:lrat-delay}]{
    \includegraphics[trim={0 1em 0 0},clip,width=\iwm\linewidth]{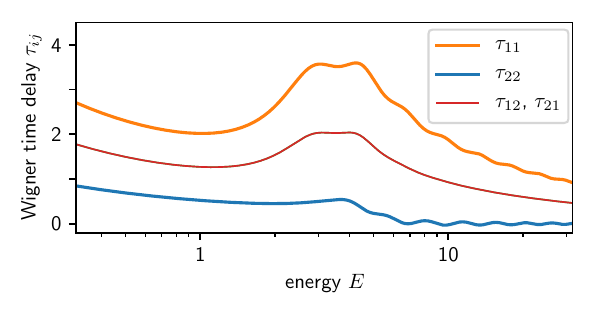}}\\
  \subfloat[\label{subfig:trat-delay}]{
    \includegraphics[trim={0 1em 0 0},clip,width=\iwm\linewidth]{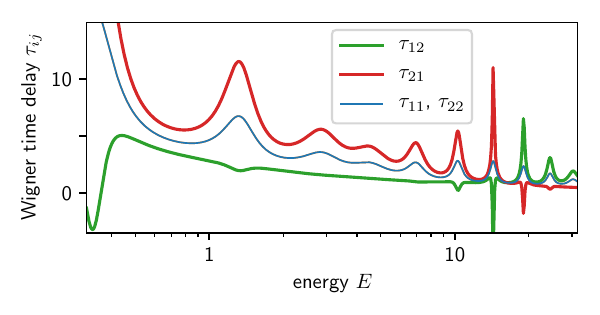}}
  \caption{\label{fig:delays}\subref{subfig:lrat-delay} Wave packet scattering delays of the asymmetric potential barrier introduced in \figref{lrat}. Reflection delays are generally different in forward and reverse direction. \subref{subfig:trat-delay} Respective plot for the asymmetric \ac{ab} interferometer introduced in \figref{trat}. For this system, the transmission delays are non-reciprocal.}
\end{figure}

The term proportional to $k^2$ in \p{ij} plays the role of a time delay in the motion of the wave packet's center. Owing to the quadratic dispersion $E=k^2$, this term corresponds to the first order of the derivative of \p{ij} with respect to energy. Figure \ref{fig:delays} shows the four resulting delays as a function of energy for the two systems introduced in the previous section. For wave packets, only the scattering phases in a certain energy window are relevant. If the packet is narrow in momentum space, this window is small and the scattering phases are linear in energy to a good approximation. ${\w{ij}}$ is then energy-independent in the relevant range. For wider wave packets, the energy dependence of the delay deforms the wave packet in addition to the broadening following from the standard ${E=k^2}$ dispersion.

\begin{figure*}
  \centering
  \begin{minipage}[t]{\rmw\linewidth}
    \subfloat[\label{subfig:lrat-packets}]{
      \includegraphics[width=\iwm\linewidth]{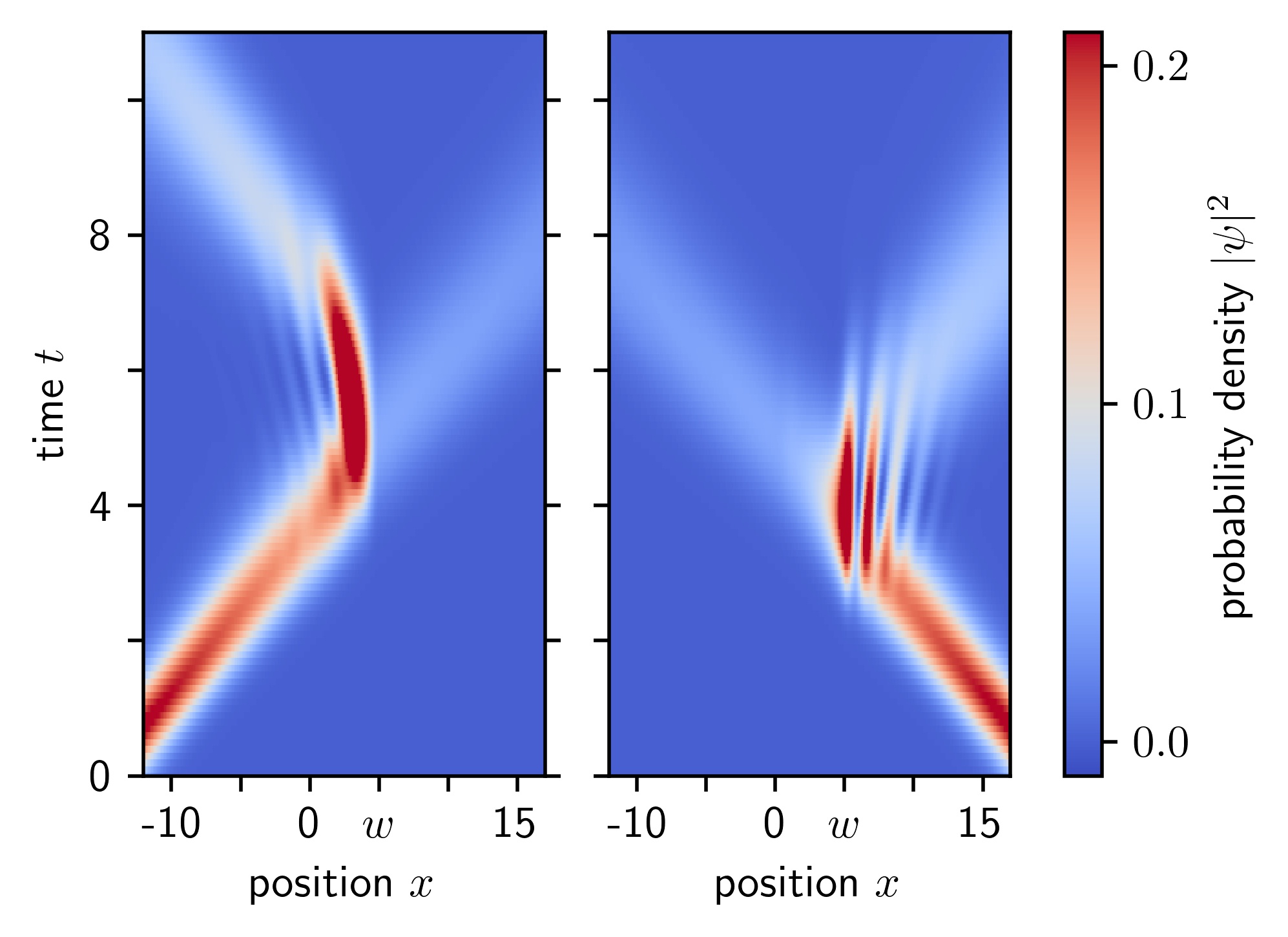}
    }
  \end{minipage}
  \begin{minipage}[t]{\rmw\linewidth}
    \subfloat[\label{subfig:trat-packets}]{
      \includegraphics[width=\iwm\linewidth]{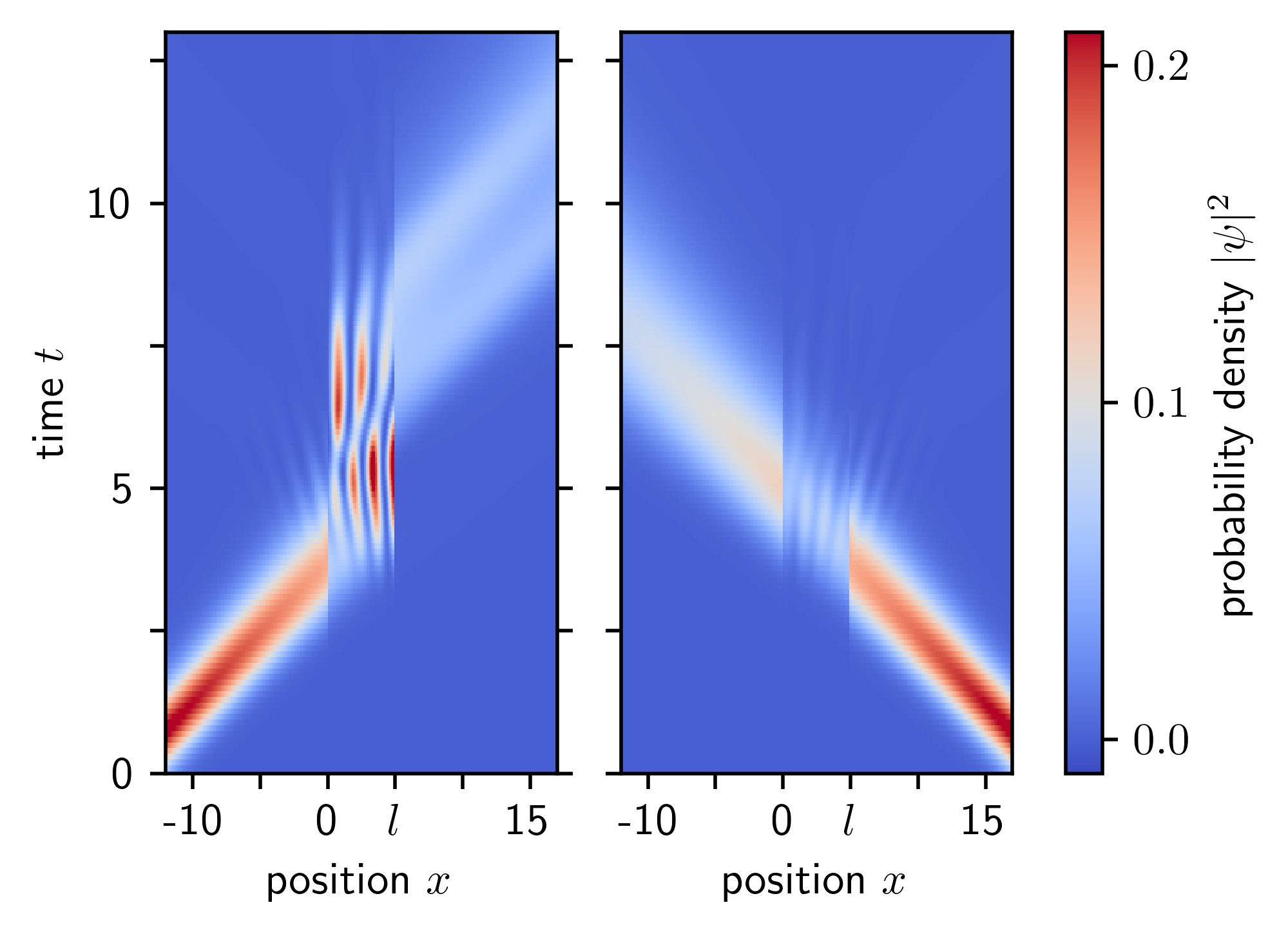}
    }
  \end{minipage}
  \caption{\label{fig:packets}\subref{subfig:lrat-packets} Time-resolved scattering of a wave packet by the asymmetric potential barrier introduced in \figref{lrat}. In the left (right) plot, the wave packet is injected from channel 1 (2). The wave packet's \acl{k0} is ${\acs{k0}=\valuekO}$, the \acl{ks} is ${\acs{ks}=\valueks}$.
          The packet position at $t=0$ is ${\acs{x0l}=\valuexOl}$ in channel 1 and ${\acs{x0r}=-\acs{x0l}+\acs{w}=-\acs{x0l}+\acs{l}=\valuexOr}$ in channel 2. \subref{subfig:trat-packets} Respective plot for the asymmetric \ac{ab} interferometer introduced in \figref{trat}. In the range $0<x<\acs{l}$,
          only the probability density of the wave function in the lower interferometer arm is shown. The wave packet parameters are equal to those used in \subref{subfig:lrat-packets}.}
\end{figure*}

Both the asymmetric potential barrier and the the asymmetric \ac{ab} interferometer are characterized by non-reciprocal wave packet scattering delays (\figref{delays}). Figure \ref{subfig:lrat-packets} shows how the non-reciprocal reflection delays ${\w{11}\neq\w{22}}$ of the asymmetric potential barrier (\figref{lrat}) affect the scattering dynamics of wave packets. A wave packet traveling towards the system from channel 2 is reflected quickly and the time spent in the scattering region is short (${3<t<5}$). The mirrored wave packet approaching from channel 1 spends a longer time inside the scattering region (${3<t<8}$) before ejection. The transmitted wave function is identical for both cases.

Figure \ref{subfig:trat-packets} shows the respective situation with the asymmetric \ac{ab} interferometer (\figref{trat}), where ${\w{21}\neq\w{12}}$. The wave packet is transmitted quickly if it approaches from channel 2. The time spent in the scattering region is short (${2.5<t<7}$). In the case where it approaches from channel 1, the wave packet experiences reflections at the beam splitters and spends a longer time (${2.5<t<10}$) in the interferometer, a behavior which was previously pointed out in \cite{nonre-inter-for}. Almost no reflection occurs because the transmission probability is exactly 1 at $\acs{k0}=\valuekO$ and close to unity in the vicinity of \acs{k0}.

Besides the time evolution of wave packets, the delays are also manifested in the eigenfunctions of the scattering problem. This relation is readily understood by taking a closer look at the eigenfunction of the asymmetric \ac{ab} interferometer (\subfigref{trat-sys}) at $E=4$, where the device is transparent. The eigenfunction with a plane wave entering only from channel 2 (``fast'' direction) consists of left-moving plane waves \emph{only}. The respective eigenfunction with plane waves entering only from channel 1 (``slow'' direction) consists of right-moving \emph{and} left-moving plane waves in the interferometer arms. The presence of both plane wave directions in the interferometer arms is necessary for the additional round trip that the wave packet takes in the interferometer in the ``slow'' direction. To be able to travel back and forth, the eigenstates making up the wave packet must contain plane waves of both propagation directions. In the ``fast'' direction, the wave packet traverses the interferometer smoothly without internal reflections. Therefore the respective eigenstates consist only of plane waves moving in one direction. Naturally, these additional plane waves contribute to the probability amplitude within the scattering region.

\begin{figure}
  \centering
  \subfloat[\label{subfig:lrat-ldos}]
        {\includegraphics[width=\iwm\linewidth]{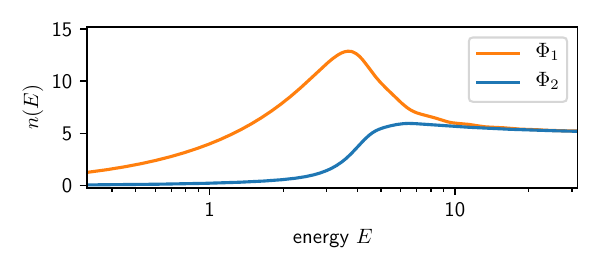}}\\
  \subfloat[\label{subfig:trat-ldos}]
        {\includegraphics[width=\iwm\linewidth]{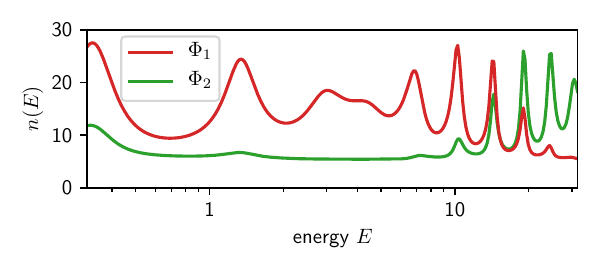}}
  \caption{\label{fig:eigen}\subref{subfig:lrat-ldos} Non-reciprocal \acl{dens} \acs{dens} of eigenfunctions of the asymmetric potential barrier introduced in \figref{lrat}. $\ac{eig}_1$ refers to eigenfunctions with an incoming plane wave from channel 1 but no incoming wave from channel 2, $\ac{eig}_2$ refers to eigenfunctions with incoming plane waves only from channel 2, not from channel 1. A comparison with \figref{delays} confirms that longer wave packet scattering delays lead to higher amplitudes of the wave function inside the scattering region. \subref{subfig:trat-ldos} Respective plot for the asymmetric \ac{ab} interferometer introduced in \figref{trat}.}
\end{figure}

This reasoning generalizes to any energy and any system: The probability density of an eigenstate in the scattering region divided by the probability current in a given channel yields a characteristic dwell time \cite{parti-densi-of-state}. We therefore expect \acm{dens}, defined as
\begin{align}
  \acs{dens}(E) = \int\limits_0^R\diff x
               \int\limits_{-\infty}^{\infty}\diff y
               \int\limits_{-\infty}^{\infty}\diff z
          \left|\eig{x,y,z}\right|^2,
\end{align}
to be associated with the wave packet scattering delays if \acmp{eig} are normalized such that the incoming wave has an amplitude of 1. Figure \ref{fig:eigen} shows $\acs{dens}(E)$ for both example systems and confirms the expected similarity with the scattering delays. The direction with the longer time delay \w{ij} also has a higher \acs{dens}. A peculiarity of this relation is that \w{ij} is defined in equation \eqnref{delay} as a derivative with respect to $E$, thus by all states around a fixed $E$. However, \acs{dens} is a property of the unique state belonging to $E$.

\section{Discussion and Conclusions}

In the case of two scattering channels the unitarity of the scattering matrix implies reciprocity for the transmission and reflection probabilities. This reciprocity cannot be broken by means of an asymmetry of the system. The wave packet scattering delays, however, are less constrained by the unitarity condition. Symmetries that are required for the probabilities are not obligatory for the scattering delays. In particular, reflection and transmission delays are non-reciprocal in two-terminal devices in two situations:
\begin{enumerate}
  \item If the scattering matrix breaks time-reversal symmetry, non-reciprocity of the transmission delays follows.
  \item If the scattering matrix is asymmetric with respect to a simultaneous reversal of time and exchange of scattering channels, the reflection delays are non-reciprocal.
\end{enumerate}

These two cases are exemplified by the asymmetric potential barrier and the asymmetric \ac{ab} interferometer, respectively.

The importance of delays for nonlinear and AC transport and the resulting breakdown of the two-terminal symmetries is well known \cite{capac-admit-and,dynam-condu-and-the,magne-asymm-of-nonli,mesos-fluct-of-nonli,viola-of-onsag-symme,magne-asymm-of-elect,wigne-time-delay-and}. However, there are fields in which the non-reciprocity presented here has not been considered so far: Coherently moving electrons can be used to carry and manipulate quantum information \cite{elect-contr-of-a}, the direction-dependent delay of wave packets opens up new possibilities for the routing and processing of such flying qubits.

An intriguing question arises if a system with non-reciprocal wave packet scattering delays is subject to decoherence. The longer the wave packet dwells in the scattering region, the stronger it is affected by decoherence, which again affects the transmission probability \cite{nonre-inter-for,quant-colla-break}. Decoherence is usually caused by inelastic or phase-breaking scattering at defects in solid state devices. A theoretical study of buckled silicene \ac{ab} rings showed that inter-valley scattering at a properly placed defect is able to convert oscillations of the wave packet scattering delay into transmission probability oscillations \cite{topol-prote-wave}. The possibility to realize non-reciprocal wave packet scattering delays and to affect the conductivity with these delays conflicts with the Onsager-Casimir conductance reciprocity \cite{recip-theor-and}. Resolving this conflict is an exciting direction for future work.

\begin{acknowledgments}

The author thanks J. Mannhart for pointing out the direction-dependent \acl{dens} and gratefully acknowledges useful discussions with him, C. Beenakker, H. Boschker, D. Braak and D. Manske and support by L. Pavka.

\end{acknowledgments}


\bibliography{main}

\begin{thebibliography}{19}%
\makeatletter
\providecommand \@ifxundefined [1]{%
 \@ifx{#1\undefined}
}%
\providecommand \@ifnum [1]{%
 \ifnum #1\expandafter \@firstoftwo
 \else \expandafter \@secondoftwo
 \fi
}%
\providecommand \@ifx [1]{%
 \ifx #1\expandafter \@firstoftwo
 \else \expandafter \@secondoftwo
 \fi
}%
\providecommand \natexlab [1]{#1}%
\providecommand \enquote  [1]{``#1''}%
\providecommand \bibnamefont  [1]{#1}%
\providecommand \bibfnamefont [1]{#1}%
\providecommand \citenamefont [1]{#1}%
\providecommand \href@noop [0]{\@secondoftwo}%
\providecommand \href [0]{\begingroup \@sanitize@url \@href}%
\providecommand \@href[1]{\@@startlink{#1}\@@href}%
\providecommand \@@href[1]{\endgroup#1\@@endlink}%
\providecommand \@sanitize@url [0]{\catcode `\\12\catcode `\$12\catcode
  `\&12\catcode `\#12\catcode `\^12\catcode `\_12\catcode `\%12\relax}%
\providecommand \@@startlink[1]{}%
\providecommand \@@endlink[0]{}%
\providecommand \url  [0]{\begingroup\@sanitize@url \@url }%
\providecommand \@url [1]{\endgroup\@href {#1}{\urlprefix }}%
\providecommand \urlprefix  [0]{URL }%
\providecommand \Eprint [0]{\href }%
\providecommand \doibase [0]{https://doi.org/}%
\providecommand \selectlanguage [0]{\@gobble}%
\providecommand \bibinfo  [0]{\@secondoftwo}%
\providecommand \bibfield  [0]{\@secondoftwo}%
\providecommand \translation [1]{[#1]}%
\providecommand \BibitemOpen [0]{}%
\providecommand \bibitemStop [0]{}%
\providecommand \bibitemNoStop [0]{.\EOS\space}%
\providecommand \EOS [0]{\spacefactor3000\relax}%
\providecommand \BibitemShut  [1]{\csname bibitem#1\endcsname}%
\let\auto@bib@innerbib\@empty
\bibitem [{\citenamefont {Wigner}(1955)}]{lower-limit-for-the}%
  \BibitemOpen
  \bibfield  {author} {\bibinfo {author} {\bibfnamefont {E.~P.}\ \bibnamefont
  {Wigner}},\ }\bibfield  {title} {\bibinfo {title} {Lower limit for the energy
  derivative of the scattering phase shift},\ }\href
  {https://doi.org/10.1103/PhysRev.98.145} {\bibfield  {journal} {\bibinfo
  {journal} {Phys. Rev.}\ }\textbf {\bibinfo {volume} {98}},\ \bibinfo {pages}
  {145} (\bibinfo {year} {1955})}\BibitemShut {NoStop}%
\bibitem [{\citenamefont {Smith}(1960)}]{lifet-matri-in-colli}%
  \BibitemOpen
  \bibfield  {author} {\bibinfo {author} {\bibfnamefont {F.~T.}\ \bibnamefont
  {Smith}},\ }\bibfield  {title} {\bibinfo {title} {Lifetime matrix in
  collision theory},\ }\href {https://doi.org/10.1103/PhysRev.118.349}
  {\bibfield  {journal} {\bibinfo  {journal} {Phys. Rev.}\ }\textbf {\bibinfo
  {volume} {118}},\ \bibinfo {pages} {349} (\bibinfo {year}
  {1960})}\BibitemShut {NoStop}%
\bibitem [{\citenamefont {Chauvat}\ \emph {et~al.}(2000)\citenamefont
  {Chauvat}, \citenamefont {Emile}, \citenamefont {Bretenaker},\ and\
  \citenamefont {Le~Floch}}]{direc-measu-of-the1}%
  \BibitemOpen
  \bibfield  {author} {\bibinfo {author} {\bibfnamefont {D.}~\bibnamefont
  {Chauvat}}, \bibinfo {author} {\bibfnamefont {O.}~\bibnamefont {Emile}},
  \bibinfo {author} {\bibfnamefont {F.}~\bibnamefont {Bretenaker}},\ and\
  \bibinfo {author} {\bibfnamefont {A.}~\bibnamefont {Le~Floch}},\ }\bibfield
  {title} {\bibinfo {title} {Direct measurement of the {W}igner delay
  associated with the {G}oos-{H}\"anchen effect},\ }\href
  {https://doi.org/10.1103/PhysRevLett.84.71} {\bibfield  {journal} {\bibinfo
  {journal} {Phys. Rev. Lett.}\ }\textbf {\bibinfo {volume} {84}},\ \bibinfo
  {pages} {71} (\bibinfo {year} {2000})}\BibitemShut {NoStop}%
\bibitem [{\citenamefont {Bourgain}\ \emph {et~al.}(2013)\citenamefont
  {Bourgain}, \citenamefont {Pellegrino}, \citenamefont {Jennewein},
  \citenamefont {Sortais},\ and\ \citenamefont
  {Browaeys}}]{direc-measu-of-the2}%
  \BibitemOpen
  \bibfield  {author} {\bibinfo {author} {\bibfnamefont {R.}~\bibnamefont
  {Bourgain}}, \bibinfo {author} {\bibfnamefont {J.}~\bibnamefont
  {Pellegrino}}, \bibinfo {author} {\bibfnamefont {S.}~\bibnamefont
  {Jennewein}}, \bibinfo {author} {\bibfnamefont {Y.~R.~P.}\ \bibnamefont
  {Sortais}},\ and\ \bibinfo {author} {\bibfnamefont {A.}~\bibnamefont
  {Browaeys}},\ }\bibfield  {title} {\bibinfo {title} {Direct measurement of
  the {W}igner time delay for the scattering of light by a single atom},\
  }\href {https://doi.org/10.1364/OL.38.001963} {\bibfield  {journal} {\bibinfo
   {journal} {Opt. Lett.}\ }\textbf {\bibinfo {volume} {38}},\ \bibinfo {pages}
  {1963} (\bibinfo {year} {2013})}\BibitemShut {NoStop}%
\bibitem [{\citenamefont {Yamahata}\ \emph {et~al.}(2019)\citenamefont
  {Yamahata}, \citenamefont {Ryu}, \citenamefont {Johnson}, \citenamefont
  {Sim}, \citenamefont {Fujiwara},\ and\ \citenamefont
  {Kataoka}}]{picos-coher-elect}%
  \BibitemOpen
  \bibfield  {author} {\bibinfo {author} {\bibfnamefont {G.}~\bibnamefont
  {Yamahata}}, \bibinfo {author} {\bibfnamefont {S.}~\bibnamefont {Ryu}},
  \bibinfo {author} {\bibfnamefont {N.}~\bibnamefont {Johnson}}, \bibinfo
  {author} {\bibfnamefont {H.-S.}\ \bibnamefont {Sim}}, \bibinfo {author}
  {\bibfnamefont {A.}~\bibnamefont {Fujiwara}},\ and\ \bibinfo {author}
  {\bibfnamefont {M.}~\bibnamefont {Kataoka}},\ }\bibfield  {title} {\bibinfo
  {title} {Picosecond coherent electron motion in a silicon single-electron
  source},\ }\href {https://doi.org/10.1038/s41565-019-0563-2} {\bibfield
  {journal} {\bibinfo  {journal} {Nat. Nanotechnol.}\ }\textbf {\bibinfo
  {volume} {14}},\ \bibinfo {pages} {1019} (\bibinfo {year}
  {2019})}\BibitemShut {NoStop}%
\bibitem [{\citenamefont {Mannhart}(2018)}]{nonre-inter-for}%
  \BibitemOpen
  \bibfield  {author} {\bibinfo {author} {\bibfnamefont {J.}~\bibnamefont
  {Mannhart}},\ }\bibfield  {title} {\bibinfo {title} {Non-reciprocal
  interferometers for matter waves},\ }\href
  {https://doi.org/10.1007/s10948-018-4637-4} {\bibfield  {journal} {\bibinfo
  {journal} {J. Supercond. Nov. Magn.}\ }\textbf {\bibinfo {volume} {31}},\
  \bibinfo {pages} {1649} (\bibinfo {year} {2018})}\BibitemShut {NoStop}%
\bibitem [{\citenamefont {Buchholz}\ \emph {et~al.}(2010)\citenamefont
  {Buchholz}, \citenamefont {Fischer}, \citenamefont {Kunze}, \citenamefont
  {Bell}, \citenamefont {Reuter},\ and\ \citenamefont
  {Wieck}}]{contr-of-the-trans}%
  \BibitemOpen
  \bibfield  {author} {\bibinfo {author} {\bibfnamefont {S.~S.}\ \bibnamefont
  {Buchholz}}, \bibinfo {author} {\bibfnamefont {S.~F.}\ \bibnamefont
  {Fischer}}, \bibinfo {author} {\bibfnamefont {U.}~\bibnamefont {Kunze}},
  \bibinfo {author} {\bibfnamefont {M.}~\bibnamefont {Bell}}, \bibinfo {author}
  {\bibfnamefont {D.}~\bibnamefont {Reuter}},\ and\ \bibinfo {author}
  {\bibfnamefont {A.~D.}\ \bibnamefont {Wieck}},\ }\bibfield  {title} {\bibinfo
  {title} {Control of the transmission phase in an asymmetric four-terminal
  {A}haronov-{B}ohm interferometer},\ }\href
  {https://doi.org/10.1103/PhysRevB.82.045432} {\bibfield  {journal} {\bibinfo
  {journal} {Phys. Rev. B}\ }\textbf {\bibinfo {volume} {82}},\ \bibinfo
  {pages} {045432} (\bibinfo {year} {2010})}\BibitemShut {NoStop}%
\bibitem [{\citenamefont {Gasparian}\ \emph {et~al.}(1996)\citenamefont
  {Gasparian}, \citenamefont {Christen},\ and\ \citenamefont
  {B\"uttiker}}]{parti-densi-of-state}%
  \BibitemOpen
  \bibfield  {author} {\bibinfo {author} {\bibfnamefont {V.}~\bibnamefont
  {Gasparian}}, \bibinfo {author} {\bibfnamefont {T.}~\bibnamefont
  {Christen}},\ and\ \bibinfo {author} {\bibfnamefont {M.}~\bibnamefont
  {B\"uttiker}},\ }\bibfield  {title} {\bibinfo {title} {Partial densities of
  states, scattering matrices, and {G}reen's functions},\ }\href
  {https://doi.org/10.1103/PhysRevA.54.4022} {\bibfield  {journal} {\bibinfo
  {journal} {Phys. Rev. A}\ }\textbf {\bibinfo {volume} {54}},\ \bibinfo
  {pages} {4022} (\bibinfo {year} {1996})}\BibitemShut {NoStop}%
\bibitem [{\citenamefont {Buttiker}(1993)}]{capac-admit-and}%
  \BibitemOpen
  \bibfield  {author} {\bibinfo {author} {\bibfnamefont {M.}~\bibnamefont
  {Buttiker}},\ }\bibfield  {title} {\bibinfo {title} {Capacitance, admittance,
  and rectification properties of small conductors},\ }\href
  {https://doi.org/10.1088/0953-8984/5/50/017} {\bibfield  {journal} {\bibinfo
  {journal} {Journal of Physics: Condensed Matter}\ }\textbf {\bibinfo {volume}
  {5}},\ \bibinfo {pages} {9361} (\bibinfo {year} {1993})}\BibitemShut
  {NoStop}%
\bibitem [{\citenamefont {B\"uttiker}\ \emph {et~al.}(1993)\citenamefont
  {B\"uttiker}, \citenamefont {Pr\^etre},\ and\ \citenamefont
  {Thomas}}]{dynam-condu-and-the}%
  \BibitemOpen
  \bibfield  {author} {\bibinfo {author} {\bibfnamefont {M.}~\bibnamefont
  {B\"uttiker}}, \bibinfo {author} {\bibfnamefont {A.}~\bibnamefont
  {Pr\^etre}},\ and\ \bibinfo {author} {\bibfnamefont {H.}~\bibnamefont
  {Thomas}},\ }\bibfield  {title} {\bibinfo {title} {Dynamic conductance and
  the scattering matrix of small conductors},\ }\href
  {https://doi.org/10.1103/PhysRevLett.70.4114} {\bibfield  {journal} {\bibinfo
   {journal} {Phys. Rev. Lett.}\ }\textbf {\bibinfo {volume} {70}},\ \bibinfo
  {pages} {4114} (\bibinfo {year} {1993})}\BibitemShut {NoStop}%
\bibitem [{\citenamefont {S\'anchez}\ and\ \citenamefont
  {B\"uttiker}(2004)}]{magne-asymm-of-nonli}%
  \BibitemOpen
  \bibfield  {author} {\bibinfo {author} {\bibfnamefont {D.}~\bibnamefont
  {S\'anchez}}\ and\ \bibinfo {author} {\bibfnamefont {M.}~\bibnamefont
  {B\"uttiker}},\ }\bibfield  {title} {\bibinfo {title} {Magnetic-field
  asymmetry of nonlinear mesoscopic transport},\ }\href
  {https://doi.org/10.1103/PhysRevLett.93.106802} {\bibfield  {journal}
  {\bibinfo  {journal} {Phys. Rev. Lett.}\ }\textbf {\bibinfo {volume} {93}},\
  \bibinfo {pages} {106802} (\bibinfo {year} {2004})}\BibitemShut {NoStop}%
\bibitem [{\citenamefont {Polianski}\ and\ \citenamefont
  {B\"uttiker}(2006)}]{mesos-fluct-of-nonli}%
  \BibitemOpen
  \bibfield  {author} {\bibinfo {author} {\bibfnamefont {M.~L.}\ \bibnamefont
  {Polianski}}\ and\ \bibinfo {author} {\bibfnamefont {M.}~\bibnamefont
  {B\"uttiker}},\ }\bibfield  {title} {\bibinfo {title} {Mesoscopic
  fluctuations of nonlinear conductance of chaotic quantum dots},\ }\href
  {https://doi.org/10.1103/PhysRevLett.96.156804} {\bibfield  {journal}
  {\bibinfo  {journal} {Phys. Rev. Lett.}\ }\textbf {\bibinfo {volume} {96}},\
  \bibinfo {pages} {156804} (\bibinfo {year} {2006})}\BibitemShut {NoStop}%
\bibitem [{\citenamefont {Szafran}\ \emph {et~al.}(2009)\citenamefont
  {Szafran}, \citenamefont {Poniedzia{\l}ek},\ and\ \citenamefont
  {Peeters}}]{viola-of-onsag-symme}%
  \BibitemOpen
  \bibfield  {author} {\bibinfo {author} {\bibfnamefont {B.}~\bibnamefont
  {Szafran}}, \bibinfo {author} {\bibfnamefont {M.~R.}\ \bibnamefont
  {Poniedzia{\l}ek}},\ and\ \bibinfo {author} {\bibfnamefont {F.~M.}\
  \bibnamefont {Peeters}},\ }\bibfield  {title} {\bibinfo {title} {Violation of
  {O}nsager symmetry for a ballistic channel {C}oulomb coupled to a quantum
  ring},\ }\href {https://doi.org/10.1209/0295-5075/87/47002} {\bibfield
  {journal} {\bibinfo  {journal} {EPL}\ }\textbf {\bibinfo {volume} {87}},\
  \bibinfo {pages} {47002} (\bibinfo {year} {2009})}\BibitemShut {NoStop}%
\bibitem [{\citenamefont {Kalina}\ \emph {et~al.}(2009)\citenamefont {Kalina},
  \citenamefont {Szafran}, \citenamefont {Bednarek},\ and\ \citenamefont
  {Peeters}}]{magne-asymm-of-elect}%
  \BibitemOpen
  \bibfield  {author} {\bibinfo {author} {\bibfnamefont {R.}~\bibnamefont
  {Kalina}}, \bibinfo {author} {\bibfnamefont {B.}~\bibnamefont {Szafran}},
  \bibinfo {author} {\bibfnamefont {S.}~\bibnamefont {Bednarek}},\ and\
  \bibinfo {author} {\bibfnamefont {F.~M.}\ \bibnamefont {Peeters}},\
  }\bibfield  {title} {\bibinfo {title} {Magnetic-field asymmetry of electron
  wave packet transmission in bent channels capacitively coupled to a metal
  gate},\ }\href {https://doi.org/10.1103/PhysRevLett.102.066807} {\bibfield
  {journal} {\bibinfo  {journal} {Phys. Rev. Lett.}\ }\textbf {\bibinfo
  {volume} {102}},\ \bibinfo {pages} {066807} (\bibinfo {year}
  {2009})}\BibitemShut {NoStop}%
\bibitem [{\citenamefont {Texier}(2016)}]{wigne-time-delay-and}%
  \BibitemOpen
  \bibfield  {author} {\bibinfo {author} {\bibfnamefont {C.}~\bibnamefont
  {Texier}},\ }\bibfield  {title} {\bibinfo {title} {{W}igner time delay and
  related concepts: Application to transport in coherent conductors},\ }\href
  {https://doi.org/https://doi.org/10.1016/j.physe.2015.09.041} {\bibfield
  {journal} {\bibinfo  {journal} {Physica E: Low Dimens. Syst. Nanostruct.}\
  }\textbf {\bibinfo {volume} {82}},\ \bibinfo {pages} {16 } (\bibinfo {year}
  {2016})}\BibitemShut {NoStop}%
\bibitem [{\citenamefont {Yamamoto}\ \emph {et~al.}(2012)\citenamefont
  {Yamamoto}, \citenamefont {Takada}, \citenamefont {Bäuerle}, \citenamefont
  {Watanabe}, \citenamefont {Wieck},\ and\ \citenamefont
  {Tarucha}}]{elect-contr-of-a}%
  \BibitemOpen
  \bibfield  {author} {\bibinfo {author} {\bibfnamefont {M.}~\bibnamefont
  {Yamamoto}}, \bibinfo {author} {\bibfnamefont {S.}~\bibnamefont {Takada}},
  \bibinfo {author} {\bibfnamefont {C.}~\bibnamefont {Bäuerle}}, \bibinfo
  {author} {\bibfnamefont {K.}~\bibnamefont {Watanabe}}, \bibinfo {author}
  {\bibfnamefont {A.~D.}\ \bibnamefont {Wieck}},\ and\ \bibinfo {author}
  {\bibfnamefont {S.}~\bibnamefont {Tarucha}},\ }\bibfield  {title} {\bibinfo
  {title} {Electrical control of a solid-state flying qubit},\ }\href
  {https://doi.org/10.1038/nnano.2012.28} {\bibfield  {journal} {\bibinfo
  {journal} {Nat. Nanotechnol.}\ }\textbf {\bibinfo {volume} {7}},\ \bibinfo
  {pages} {247} (\bibinfo {year} {2012})}\BibitemShut {NoStop}%
\bibitem [{\citenamefont {Bredol}\ \emph {et~al.}(2019)\citenamefont {Bredol},
  \citenamefont {Boschker}, \citenamefont {Braak},\ and\ \citenamefont
  {Mannhart}}]{quant-colla-break}%
  \BibitemOpen
  \bibfield  {author} {\bibinfo {author} {\bibfnamefont {P.}~\bibnamefont
  {Bredol}}, \bibinfo {author} {\bibfnamefont {H.}~\bibnamefont {Boschker}},
  \bibinfo {author} {\bibfnamefont {D.}~\bibnamefont {Braak}},\ and\ \bibinfo
  {author} {\bibfnamefont {J.}~\bibnamefont {Mannhart}},\ }\href@noop {}
  {\bibinfo {title} {Quantum collapses break reciprocity in matter transport}}
  (\bibinfo {year} {2019}),\ \Eprint {https://arxiv.org/abs/1912.11948}
  {arXiv:1912.11948 [cond-mat.mes-hall]} \BibitemShut {NoStop}%
\bibitem [{\citenamefont {Szafran}\ \emph {et~al.}(2019)\citenamefont
  {Szafran}, \citenamefont {Rzeszotarski},\ and\ \citenamefont
  {Mrenca-Kolasinska}}]{topol-prote-wave}%
  \BibitemOpen
  \bibfield  {author} {\bibinfo {author} {\bibfnamefont {B.}~\bibnamefont
  {Szafran}}, \bibinfo {author} {\bibfnamefont {B.}~\bibnamefont
  {Rzeszotarski}},\ and\ \bibinfo {author} {\bibfnamefont {A.}~\bibnamefont
  {Mrenca-Kolasinska}},\ }\bibfield  {title} {\bibinfo {title} {Topologically
  protected wave packets and quantum rings in silicene},\ }\href
  {https://doi.org/10.1103/PhysRevB.100.085306} {\bibfield  {journal} {\bibinfo
   {journal} {Phys. Rev. B}\ }\textbf {\bibinfo {volume} {100}},\ \bibinfo
  {pages} {085306} (\bibinfo {year} {2019})}\BibitemShut {NoStop}%
\bibitem [{\citenamefont {{Casimir}}(1963)}]{recip-theor-and}%
  \BibitemOpen
  \bibfield  {author} {\bibinfo {author} {\bibfnamefont {H.~B.~G.}\
  \bibnamefont {{Casimir}}},\ }\bibfield  {title} {\bibinfo {title}
  {Reciprocity theorems and irreversible processes},\ }\href
  {https://doi.org/10.1109/PROC.1963.2627} {\bibfield  {journal} {\bibinfo
  {journal} {Proceedings of the IEEE}\ }\textbf {\bibinfo {volume} {51}},\
  \bibinfo {pages} {1570} (\bibinfo {year} {1963})}\BibitemShut {NoStop}%
\end{thebibliography}%

\end{document}